\documentclass[twocolumn,showpacs,preprintnumbers,amsmath,amssymb,prb,superscriptaddress]{revtex4}

\usepackage{graphicx}
\usepackage{color}
\usepackage{epsfig}
\usepackage{overpic}
\usepackage[colorlinks]{hyperref}

\begin{document}
 
\title{Transport properties of hybrid topological superconductor devices in contact with an environment}
\author{A. Komnik}
\affiliation{Institut f\"ur Theoretische Physik, Universit\"at Heidelberg, Philosophenweg 12, D-69120 Heidelberg, Germany}

\date{\today}

\begin{abstract}

We investigate the transport properties of different realizations of one-dimensional quantum wires coupled to a number of external electrodes
in terms of the full counting statistics. Focusing on the set-ups in which edge states of Majorana type are realizable we analyze the effects of the exchange of charge carriers with the environment. We find a very strong suppression of the non-local current correlations, which decay  exponentially with the coupling strength and the wire length. The local properties such as transport currents and current correlations measured at individual contacts between the wire and external electrodes turn out to develop asymmetry when energy-resolved. We provide a number of exact analytical solutions and discuss experimental implications of our findings.

\end{abstract}


\maketitle

\section{Introduction}
\label{sec:intro}

The physics of topological insulators and superconductors is one of the most active fields of research. Especially for future applications in information processing technology it is important to understand the transport properties of nano- and microscopic structures based on low-dimensional systems with topologically non-trivial band structures.\cite{QuantumComputingReview2008} Here a special role play (quasi)-one-dimensional electronic systems subject to triplet superconducting pairing, also known under the name of Kitaev chains.\cite{Kitaev2001} Their most important feature is the presence of highly non-trivial end states, which can be shown to be a realization of the famous Majorana fermions. Right now we witness a tremendous progress in manufacturing of such devices and the existence of such Majorana states have been shown in several works.\cite{Mourik25052012,MajoranaExpHeiblum2012,MajoranaMarcus2013}

The prototype system of this kind is an appropriately gated quantum wire with very strong spin-orbit coupling (on the basis of InSb, InAs and related compounds), which is subject to magnetic field and which is in contact with a conventional superconductor.\cite{Lutchyn2010,OregRefaelvonOppen2010,StanescuDasSarma2011,Mourik25052012,PhysRevB.87.024515}  Such kind of systems requires a very high degree of control. On the other hand there are also proposals using chains of quantum dots, which can be controlled with very high degree of accuracy.\cite{SauDasSarma2012,FulgaAkhmerov2013}  The quantum wires are usually modelled by continuum versions of field theory (see e.~g. Ref.~[\onlinecite{KueiSunShah_2015}]) while the quantum dot arrays intrinsically require lattice models. They are much more difficult to handle especially in geometries with open boundaries, which are necessary for the topologically non-trivial states to form. Thus
one is confronted with a solution of boundary value problems for finite systems, which can often become rather complex and intractable. 
 A very efficient tool for modelling of such systems is the recurrence relation technique which goes  back to Refs.~[\onlinecite{Haydock1,MacKinnon:1985rt}]. It allows for working with lattice models directly and results in very fast converging numerics. It became the method of choice in the modelling of systems with topologically non-trivial band structures. However, exact solutions are few and far between especially for systems in non-equilibrium. One of the goals of the present work is to show the general solution strategy and to present a number of solutions in different set-ups.

There are several aspects which have not yet been considered in full detail and which, on the other hand, are very important in realistic systems. One of them is the contact of the quantum wire with the environment. While the gating is usually appropriately taken into account and its effects are well understood,\cite{VuikWimmer2016} a genuine leakage or particle exchange with the substrate is rarely considered. The analysis of influence of these processes is the goal of our study. We discuss it in a simple geometry, in which a quantum wire is contacted at its ends by metallic electrodes, which induce finite electric current through it. The wire is modelled by a finite size lattice, in which each of its sites is coupled to an electron bath via tunnelling with strength $\delta$. It turns out that $\delta\neq 0$ profoundly alters the behaviour of systems in question, significantly limiting the observability of such non-trivial effects as Majorana edge states. In order to include the coupling to the environment we have developed an extension of the recurrence relation approach to non-equilibrium Keldysh Green's functions (GF). Furthermore, we demonstrate that a dedicated chain contraction procedure helps to significantly reduce the number of computational steps necessary to access transport properties. 
Using these techniques we are able to produce compact expressions for the cumulant generating function (CGF) for cumulants of charge transferred through the wire. We also provide full analytical solutions whenever possible.

The structure of the paper is as follows: In the next Section we outline the formalism, which is used for subsequent calculations. We show that it is possible to write down a generalized Hamiltonian, which is suitable for modelling of a wide class of quantum wires contacted by transport electrodes as well as an environment. After that we perform the chain contraction procedure and present the main result -- the expression for the CGF. In Section III we apply the developed method to three different systems: (a) to a non-interacting wire; (b) to a Kitaev chain; and (c) to a quantum wire with strong spin-orbit interaction (SOI) in contact with a conventional superconductor. It is shown, that situations (a) and (b) allow for an exact analytical solution of the problems. We thoroughly analyze the conditions under which the Majorana endstates are visible in current correlations and make predictions for the future experiments. The Conclusions section summarizes our results.

\section{Formalism}

In the course of our study we shall consider one-dimensional fermionic systems with internal degrees of freedom, the most obvious of which is the spin. We model them by versions of the tight-binding Hamiltonian containing $N$ sites. It is formulated best in the \emph{composite field}  notation
 ${\bf c}_i^\dag = (c_{i \uparrow}^\dag, c_{i \uparrow}, c_{i \downarrow}^\dag, c_{i \downarrow})$, where we anticipate the presence of superconducting (SC) couplings by combining creation and annihilation operators to a single composite field, 
\begin{eqnarray}   \label{KitaevDvector}
 H_0 = \sum_{i=1}^{N} {\bf c}_i^\dag \boldsymbol{\epsilon} {\bf c}_i +
 \sum_{i=1}^{N-1} \left( {\bf c}_i^\dag \boldsymbol{\Gamma} {\bf c}_{i+1} +
  {\bf c}_{i+1}^\dag \boldsymbol{\Gamma}^\dag {\bf c}_{i} 
  \right) \, .
\end{eqnarray}
Here $\boldsymbol{\epsilon}$ describes the on-site part of the Hamiltonian and $\boldsymbol{\Gamma}$ are the matrices responsible for the inter-site coupling. 
For example in a simple tight binding chain, in which the hopping amplitude between adjacent sites is some constant $\gamma$ we obtain 
\begin{eqnarray}    \label{TBexample}
 {\bf \Gamma} = \mbox{diag}(\gamma, - \gamma, \gamma, - \gamma) = 
\gamma \, \mbox{diag}(\sigma_z, \sigma_z) \, ,
\end{eqnarray}
where $\sigma_z= \mbox{diag}(1,-1)$ is a Pauli matrix.
A particle transport through such a chain is accomplished by coupling one or several of its sites by metallic electrodes via tunnelling. In the simplest two-electrode configuration it is allowed to tunnel between the site $j=1$ and the left electrode and between the site $j=N$ and the right electrode, so that 
\begin{eqnarray}
 H = H_0 + H_L + H_R + {\bf L}^\dag \boldsymbol{\Lambda}_L {\bf c}_1  + 
 {\bf R}^\dag \boldsymbol{\Lambda}_R {\bf c}_N + \text{H.c.} \, , 
\end{eqnarray}
where $H_{L,R}$ describe the electrode degrees of freedom with ${\bf R,L}$ being the respective composite fields, and $\boldsymbol{\Lambda}_{L,R}$ are the coupling matrices, which are in general different from intra-chain $\boldsymbol{\Gamma}$. 

In realistic set-ups the wire can exchange particles with the environment, for instance with the substrate. One can very efficiently model that by coupling the individual sites of the lattice to a fermionic continuum with a fixed chemical potential. Thus the full Hamiltonian is given by
\begin{eqnarray}
 H_\text{full} = H + H_\text{env} \, , 
\end{eqnarray}
where 
\begin{eqnarray}
 H_\text{env} = \sum_{i=1}^N (\gamma_i \, {\bf c}_i^\dag \, {\bf \Gamma}_i \, \boldsymbol{\psi}_i  + \text{H.c.}) + H_\psi \, . 
\end{eqnarray}
Here ${\bf \Gamma}_i = \gamma_i \, \mbox{diag}(\sigma_z, \sigma_z)$, where
 $\gamma_i$ are tunnelling amplitudes between the environment and the respective lattice site. In most of our discussions later we consider them to be equal to $\gamma_0$. 
$\boldsymbol{\psi}_i$ denotes the composite field of the continuum.
As far as its properties are concerned the only necessary parameters are the density of states $\rho_0$ and its chemical potential, which we assume to be fixed at zero throughout. In this framework the only important energy scale is $\delta = \rho_0 \gamma_0^2$, which is equal to the reciprocal average lifetime of an isolated lattice site due to leakage into the continuum.

Physical properties of the above systems are best accessed by functional integration in Keldysh space as we are dealing with inherently non-equilibrium systems. The second reason is that the primary quantity of our interest is the cumulant generating function (CGF) of the charge transfer statistics, which is very conveniently calculated in this framework. 
The CGF is essentially Keldysh partition function and can be written down as a Grassmannian integral
\begin{eqnarray}    \label{funInt0}
Z = \int {\cal D}[L^\dag,R^\dag, L,R]  \prod_{i=1}^N {\cal D}[c^\dag_i,c_i] \, \exp( i S) \, ,
\end{eqnarray}
in which the environment is already integrated out and where the action is given by
\begin{eqnarray}
S &=& \sum_{\alpha=L,R} ( S_\alpha + T_\alpha ) + S_N \, .
\\
S_N &=& \sum_{i=1}^{N} \int_C d t d t' c_i^\dag(t) D_0^{-1}(t-t') c_i (t')
 \\  \nonumber
 &+& \int_C d t
 \sum_{i=1}^{N-1} [ c_{i}^\dag(t) \Lambda c_{i+1}(t) + c^\dag_{i+1}(t) \Lambda^\dag c_{i}(t) ] \, ,
\end{eqnarray}
where $c_i(t)$, $L(t)$, $R(t)$ and their adjoint are the extensions of the above composite fields to the Keldysh space so that by abuse of notation 
\begin{eqnarray}                 \label{LongC}
 c_i^\dag =  (c_{i \uparrow -}^\dag, c_{i \uparrow +}^\dag, c_{i \uparrow -}, c_{i \uparrow +}, c_{i \downarrow -}^\dag, c_{i \downarrow +}^\dag, c_{i \downarrow-}, c_{i \downarrow +}) \, . 
\end{eqnarray}
We call them \emph{composite action fields}. 
The additional index $\pm$ denotes the field on the forward/backward subcontour of the Keldysh path $C$. Here $\Lambda$ denotes matrices acting in the composite action field space, which couples the composite fields on adjacent sites. They can be straightforwardly constructed from ${\bf \Gamma}$s of \eqref{KitaevDvector}. For instance, in the case of a simple tight binding chain with ${\bf \Gamma}$ given in \eqref{TBexample} one obtains
$\Lambda= \gamma \, $diag$(\sigma_z, - \sigma_z, \sigma_z, - \sigma_z)$.
$D_0(t)$ stands for the Keldysh GF matrices of the individual sites coupled to the continuum but decoupled from its counterpart within the lattice. It can be easily computed with the knowledge of the respective $\boldsymbol{\epsilon}$ and the coupling to the substrate $\delta$. 
$S_\alpha$ denote the action of the particles in the electrodes.
In the overwhelming number of experimental set-ups the leads can be assumed to be metallic with nearly free electrons. Hence 
\begin{eqnarray}   \label{}
 S_\alpha = \int_C d t d t' \, \alpha^\dag (t) G_\alpha^{-1} (t-t') \alpha (t') \, .
\end{eqnarray}
Here $G_\alpha$ is the local single particle GF matrix which has the form 
\begin{eqnarray}   \label{}
 G_\alpha^{-1}(\omega) = \mbox{diag} \left(
 g_{\alpha \uparrow}^{-1}(\omega), 
  \widetilde{g}_{\alpha \uparrow}^{-1}(\omega) ,
  g_{\alpha \downarrow}^{-1}(\omega), 
  \widetilde{g}_{\alpha \downarrow}^{-1}(\omega) \right) \, ,
\end{eqnarray}
with\cite{CaroliNozieresI}
\begin{eqnarray}     \label{leadGF}
  g_{\alpha \sigma}(\omega) = i \rho_\alpha(\omega) 
  \left(
\begin{array}{cc}
 n_\alpha - 1/2 & n_\alpha \\
 n_\alpha - 1 & n_\alpha - 1/2
\end{array}
\right) \, .
\end{eqnarray}
$\rho_\alpha(\omega)$ is the energy-dependent density of states in the respective electrode and $n_\alpha$ describes the Fermi distribution function with chemical potential $V_\alpha$. $\widetilde{g}_{\alpha \sigma}$ is the GF of the holes. It is obtained from \eqref{leadGF} by the transformation $n_\alpha \to 1 - n_\alpha$.   

The coupling between the wire and the electrodes is accomplished by
\begin{eqnarray}
 \sum_\alpha T_\alpha = \int_C d t \,  L^\dag(t) \Lambda_L c_1(t)  + 
 R^\dag(t) \Lambda_R c_N(t)  + \text{adj} 
\end{eqnarray}
As was suggested in [\onlinecite{LevitovReznikov2004}] and successfully applied to tunnelling systems in Refs.~[\onlinecite{TowardsFCS, FCSKondo2005}], the CGF can be computed by introduction of the counting field $\lambda$ as a phase at each tunnelling vertex. Although it is constant within each of the Keldysh subcontours, it carries different signs on them. In a very good approximation we can assume the tunnelling processes to take place instantaneously, then
\begin{eqnarray}   \label{}
\Lambda_\alpha &=& \gamma_\alpha
 \\ \nonumber &\times&
  \left(
  e^{i \lambda_{\alpha \uparrow}}, - e^{-i \lambda_{\alpha \uparrow}}, 1, -1, 
   e^{i \lambda_{\alpha \downarrow}},  -e^{-i \lambda_{\alpha \downarrow}}, 1, -1
 \right)
\end{eqnarray}
where $\gamma_\alpha$ is the tunnelling matrix element and $\lambda_{\alpha, \sigma}$ is the respective counting field. 

The above action is quadratic in fermion fields and thus the functional integral can be calculated by elementary methods. The most economical strategy is to integrate out the bulk sites of the wire first. To that end we split Eq.~\eqref{funInt0} into two integrals: (i) one over the outmost wire sites and electrode fields; (ii) the one over the bulk sites of the wire, which is given by
\begin{eqnarray}    \label{funIntbulk}
Z' &=& \int  \prod_{i=2}^{N-1} {\cal D}[c^\dag_i,c_i] \, \exp(i S_2)
 \\ \nonumber
&=& \exp\left[ 
i \int_C d t d t' \, (c_1^\dag, c_N^\dag)(t) {\cal A}_2(t-t') \left( \begin{array}{c}
c_1 \\ c_N
\end{array} \right)(t')
\right]
 \, ,
\end{eqnarray}
with the kernel [we explain how to compute it after Eq.~\eqref{iter1}]
\begin{eqnarray}                         \label{Adef}
 {\cal A}_2  = \left(
\begin{array}{cc}
{D}_{N-2}^{-1} & \Lambda_{N-2} \\
{\Lambda}^\dag_{N-2} &  {D}_{N-2}^{-1} 
 \end{array}
\right) \, .
\end{eqnarray}
Its meaning is rather simple: its inverse yields the matrix
\begin{eqnarray}    \label{invA}
  {\cal A}_2^{-1} = \left(
\begin{array}{cc}
 D_{11} & D_{1N} \\
 D_{N1} & D_{NN}
\end{array}
\right) \, , 
\end{eqnarray}
where $D_{ij}(t) = - i \langle T_C c_i(t) c^\dag_j(0) \rangle$ are the Keldysh GFs between the respective sites ($T_C$ denotes the contour ordering operator). The diagonal components of the last matrix are thus the local GFs of the outmost chain sites whereas the off-diagonal ones correspond to the particle propagation between the wire ends. It is possible to derive explicit expressions yielding the elements of  \eqref{Adef} in the closed form. 
There are two ways to do that: by computing the GFs \eqref{invA} and inverting the matrix, or by successively integrating out the bulk sites until only the fields for the outmost ones remain in the action. We choose the latter procedure and write down the action kernel in \eqref{funIntbulk} in an extended notation, in which it is quadratic in the composite fields obtained out of those in \eqref{LongC} by concatenating them in the order they occur along the lattice. For the partition function we then can write a functional integral
\begin{eqnarray}   \label{SNzed}
 Z' = \int {\cal D} [c^\dag, c] \, e^{i S_N} \, ,
\end{eqnarray}
where the action $S_N$ is 
\begin{eqnarray}   \label{startAction}
 S_N = \int_C d t d t' c^\dag (t) {\cal A}_N(t-t') c(t') \, .
\end{eqnarray}
Here we use a new object
\begin{eqnarray}     \label{SuperC}
c^\dag = (c_1^\dag, c_2^\dag, \dots , c_N^\dag)
\end{eqnarray}
which is the concatenation of the original composite action fields mentioned above. The action kernel is a large matrix of the form
\begin{eqnarray}        \label{iter0}
{\cal A}_N = \left(
\begin{array}{cccccc}
{D}_0^{-1} & \Lambda & 0 & \dots & \dots & 0 \\
\Lambda^\dag &  {D}_0^{-1} & \Lambda & 0  & \dots & 0 \\
0 & \Lambda^\dag &  {D}_0^{-1} & \Lambda  & \dots & 0 \\
0 & 0 & \Lambda^\dag &  {D}_0^{-1}  & \dots & 0 \\
\dots  & \dots  & \dots  & \dots  & \dots  & \Lambda \\
0 & \dots & 0   & \dots&  \Lambda^\dag &  {D}_0^{-1}
\end{array}
\right) \, . 
\end{eqnarray}
The subscript $N$ refers to the number of sites, composite fields of which enter Eq.~\eqref{SuperC}. In order to obtain the representation \eqref{funIntbulk} we integrate out the composite fields with indices $j=2, \dots, N-1$ one after the other and in this order in a procedure we call \emph{chain contraction}. 
After integrating out $n$ sites (composite fields) we have instead of  \eqref{SNzed}
\begin{eqnarray}   \label{}
 Z' = \int {\cal D} [\tilde{c}^\dag, \tilde{c}] \, e^{i S_{N-n}} \, ,
\end{eqnarray}
where $\tilde{c}^\dag = (c_1^\dag, c_{n+2}^\dag, \dots , c_N^\dag)$ has a reduced dimension, and
\begin{eqnarray}   \label{startAction}
 S_{N-n} = \int_C d t d t' \, \tilde{c}^\dag (t) {\cal A}_{N-n} (t-t') \tilde{c}(t') \, ,
\end{eqnarray}
with
\begin{eqnarray}        \label{iter1}
{\cal A}_{N-n} = \left(
\begin{array}{cccccc}
{D}_n^{-1} & \Lambda_n & 0 & \dots & \dots & 0 \\
\Lambda^\dag_n &  {F}_n^{-1} & \Lambda & 0  & \dots & 0 \\
0 & \Lambda^\dag &  {D}_0^{-1} & \Lambda  & \dots & 0 \\
0 & 0 & \Lambda^\dag &  {D}_0^{-1}  & \dots & 0 \\
\dots  & \dots  & \dots  & \dots  & \dots  & \Lambda \\
0 & \dots & 0   & \dots&  \Lambda^\dag &  {D}_0^{-1}
\end{array}
\right) \, . 
\end{eqnarray}
 ${D}_n$, $\Lambda_n$ and ${F}_n$ are matrices changing during the chain contraction. 
Before we start it (for $n=0$) $D_n = F_n = D_0$ is obviously the GF of a bare uncoupled site and $\Lambda_n = \Lambda$. It turns out that it is possible to derive recurrence relations for $D_n$, $F_n$ and $\Lambda_n$. To accomplish that we start with 
\eqref{iter1}, which is the action kernel after $n$ steps of chain contraction. 
In the next one the action matrix becomes 
\begin{eqnarray}        \label{iter2}
 \left(
\begin{array}{ccccc}
{D}_n^{-1} -  \Lambda^\dag_n  {F}_n \Lambda_n & -\Lambda_n { F}_n \Lambda & 0 & \dots & \dots  \\
-\Lambda^\dag {F}_n \Lambda_n^\dag &  {D}_0^{-1}  - \Lambda^\dag { F}_n \Lambda  & \Lambda & 0  & \dots  \\
0 & \Lambda^\dag &  {D}_0^{-1} & \Lambda  & \dots \\
0 & 0 & \Lambda^\dag &  {D}_0^{-1}  & \dots  \\
\dots  & \dots  & \dots  & \dots  & \dots   \\
\end{array}
\right) \, . 
\end{eqnarray}
In this way we obtain the following recurrence relations: 
\begin{eqnarray}   \label{rec_system}
  F^{-1}_{n+1} &=&  {D}_0^{-1}  - \Lambda^\dag { F}_n \Lambda \, , 
 \\   \label{rec_system1}
  {D}^{-1}_{n+1} &=& {D}_n^{-1} -  \Lambda^\dag_n  {F}_n \Lambda_n
   \, , 
    \\ \label{rec_system2}
   \Lambda_{n+1} &=& -\Lambda_n { F}_n \Lambda \, ,  \, \, \,
 \Lambda_n^\dag = -\Lambda^\dag {F}_n \Lambda_n^\dag \, .
\end{eqnarray}
After $N-2$ steps we arrive at \eqref{Adef}. Needless to say, after this last step $F_{N-2} = D_{N-2}$. Contracting even further we shall obtain the effective action of the outmost chain site. The respective action kernel is the inverse GF at the end is 
\begin{eqnarray}    \label{enddef}
F_{N-1}={\cal G}_\text{end}(N) \, .
\end{eqnarray} 
It is obvious, that once one has solved the recurrence relation \eqref{rec_system} the solutions of other three easily follow. It turns out that \eqref{rec_system} is a version of a matrix valued continued fraction\cite{Zhao2003136} of a finite length and is 
 indeed formally analytically solvable for any finite $n$ and any coupling matrices $\Lambda$.  
We show how to do that in the simpler case of $\Lambda=\Lambda^\dag$ in Section \ref{Nonintchain}, while the solution for a situation of a generic $\Lambda$ is discussed in Section \ref{Kitaev_chain}.

Now the partition function  \eqref{funInt0} is simplified to 
\begin{eqnarray}    \label{funInt3}
Z &=& \int {\cal D}[L^\dag,R^\dag, L,R] \, {\cal D}[c^\dag_1,c_1,c^\dag_N,c_N] \, 
\nonumber \\ 
&\times&
Z' \exp \left[
i \sum_{\alpha=L,R} ( S_\alpha + T_\alpha ) \right] \, .
\end{eqnarray}
It is reasonable to assume that the density of states in the electrodes is only weakly energy dependent at least in the nearest vicinity of the Fermi edge, thus we simplify it by a constant $\rho_\alpha (\omega) \approx \rho_\alpha$.\footnote{This is an inessential simplification. Any other more realistic band structure can be considered as well.} Integrating over the electrode degrees of freedom we find
\begin{eqnarray}    \label{funInt3}
Z &=& \int {\cal D}[c^\dag_1,c_1,c^\dag_N,c_N] \,
\\ \nonumber 
&\times& \exp\left[ 
i \int_C d t d t' \, (c_1^\dag, c_N^\dag)(t) {\cal A'}(t-t') \left( \begin{array}{c}
c_1 \\ c_N
\end{array} \right)(t')
\right]
\, ,
\end{eqnarray}
with
\begin{eqnarray}   \label{FCS1}
{\cal A'}  = \left(
\begin{array}{cc}
{D}_{N-2}^{-1} - \Lambda_L^\dag G_L \Lambda_L & \Lambda_{N-2} \\
{\Lambda}^\dag_{N-2} &  {D}_{N-2}^{-1}  - \Lambda_R^\dag G_R \Lambda_R
 \end{array}
\right) \, .
\end{eqnarray}
The remaining functional integral can be expressed in terms of the determinant of this matrix, 
\begin{eqnarray}   \label{FCS2}
 \ln Z(\lambda_L,\lambda_R) = {\cal T} \int \frac{d \omega}{2 \pi} \, \ln \mbox{det} \,  {\cal A'} = F(\lambda_L,\lambda_R) \, .
\end{eqnarray}
This is precisely the CGF $F(\lambda_L,\lambda_R) $ of charge transfer during very long waiting time ${\cal T} \to \infty$. Any conceivable irreducible moments (cumulants) of the charges transferred through the contacts during ${\cal T}$ can be computed using the relation
\begin{eqnarray}   \label{cumulants}
\langle \delta^n Q_L \delta^m Q_R \rangle = (-i)^{n + m} \frac{\partial^{n+m} F(\lambda_L,\lambda_R)}{\partial^n \lambda_L \, \partial^m \lambda_R} \Big|_{\lambda_L,\lambda_R=0} \, .
\end{eqnarray}
In particular, the plain average currents through the contacts are just $ I_\alpha = {\cal T}^{-1}  \langle \delta Q_\alpha \rangle$, while the noise power is $ S_0 = {\cal T}^{-1} \langle \delta^2 Q_\alpha \rangle $. Another interesting quantity is the current cross-correlation -- it is proportional to\cite{Weithofer2014,PhysRevB.90.195404,StrubiBelzig2011} 
 \begin{eqnarray}   \label{CurrentCross}
 S_{LR} = \langle \delta I_L \delta I_R \rangle = {\cal T}^{-1} \langle \delta Q_L \delta Q_R \rangle \, ,
\end{eqnarray}
where $I_{L,R}$ are the currents through the respective contact. 
These are the quantities of our primary interest and we address their properties in the next section. 
We focus on the energy-resolved data, which are obtained by leaving out the integration over energy in Eq.~\eqref{FCS2}.  At zero temperature it is largely equivalent to the derivatives of the cumulants with respect to the applied voltage, e.~g. the energy resolved current is then equal to the differential conductance.\cite{Weithofer2014}

\section{Results}

In this section we apply the presented method and consider different systems with different kinds of interactions within the wire as well as between the wire and the substrate.

\subsection{Non-interacting chain}
\label{Nonintchain}

First we do not subject the wire to magnetic field and abstain from including superconductivity, so that we work with a simple spinless tight-binding chain
\begin{eqnarray}      \label{H0x}
 H_\text{tb} = \sum_{i=1}^N \epsilon \, c_i^\dag c_i + \sum_{i=1}^{N-1} \gamma ( c_{i}^\dag c_{i+1} + c^\dag_{i+1} c_{i} ) \, ,
\end{eqnarray} 
where we assume all on-site energies to be equal to $\epsilon$. 
Then the composite action field \eqref{LongC} only contains two components, $c^\dag_i = (c_{i  -}^\dag, c_{i +}^\dag)$. The GF of an uncoupled site with energy $\epsilon$ is $D_0(\omega) = \mbox{diag} ((\omega -\epsilon)^{-1},  -(\omega -\epsilon)^{-1})$. This kind of GF does not carry any information about the filling fraction of the site. Moreover, its spectral function is a delta-function shaped peak describing a state with infinite lifetime. A more realistic is a situation, in which every single site of the chain has finite width in the energy domain and hence a finite lifetime $1/\delta$. After integrating out the environment we then obtain 
\begin{eqnarray}   \label{}
  {D}_0^{-1}(\omega) = 
   \left(
 \begin{array}{cc}
 \omega + i \delta (2 n_F - 1) & - i \delta 2 n_F \\
 - i \delta 2 (n_F - 1) &  -\omega + i \delta (2 n_F - 1) 
 \end{array}
 \right) \, ,
\end{eqnarray}
where $n_F$ is a Fermi distribution function with the edge at $\omega=0$. We also set $\epsilon=0$
(this choice enforces half-filling of the individual lattice sites). The coupling matrix is obviously a 2$\times$2 matrix too and can be written down as $\Lambda = \gamma \, \sigma_z$. According to \eqref{rec_system} we have to solve the recurrence relation
\begin{eqnarray}   \label{}
  F^{-1}_{n+1} &=&  {D}_0^{-1}  - \gamma^2 \sigma_z { F}_n \sigma_z \, .
\end{eqnarray}
We recall the factorization property
\begin{eqnarray}   \label{}
  \sigma_z = i \sigma_y \sigma_x = 
  \left(
\begin{array}{cc}
0 & 1 \\
-1 & 0
\end{array}
\right) \left(
\begin{array}{cc}
0 & 1 \\
1 & 0
\end{array}
\right)
\end{eqnarray}
and define a new unknown $\widetilde{F}_n =  \sigma_x\, {F}_{n} \,  i \sigma_y$, the equation for which does not involve matrix products any more,
 \begin{eqnarray}   \label{}
  \widetilde{F}^{-1}_{n+1} &=&  \widetilde{D}_0^{-1}  - \gamma^2 \widetilde{F}_n  \, .
\end{eqnarray}
$\widetilde{D}_0$ is an intrinsically regular matrix and as such can be diagonalized via
matrix ${U}$: ${A}_0 = {U}^{-1} \widetilde{D}_0 {U}$. Interestingly $A_0 = \mbox{diag}(G_0^{R} ,  G_0^{A} )$, where $G_0^{R,A} = 1/(\omega \pm i \delta)$ are the retarded/advanced GFs of an individual level with a width $\delta$. The transformation matrices are
\begin{eqnarray}      \label{H0}
{U} =
 \left(
\begin{array}{cc}
 \frac{n_F-1}{n_F} & 1 \\
 1 & 1 
\end{array}
\right) \, , \, \, \, \, 
{U}^{-1} =
 \left(
\begin{array}{cc}
-n_F & n_F \\
 n_F & 1-n_F 
\end{array}
\right) \, .
\end{eqnarray}
The recursion relation to solve is now
\begin{eqnarray}   \label{}
  {U}^{-1}  \widetilde{F}^{-1}_{n+1} {U}  = {A}^{-1}_0  - \gamma^2 \,  {U}^{-1} \widetilde{F}_{n}  {U} \, . 
\end{eqnarray}
As the source term of this relation is diagonal it is reasonable to assume that $ {U}^{-1}  \widetilde{F}^{-1}_{n} {U} = B_n^{-1}$ are diagonal too. We have performed numerical checks, which confirm this conjecture. Additional argument in its favour is the fact that it is definitely true in the relevant limit $n \to \infty$. Thus we are confronted with two scalar recurrence relations
\begin{eqnarray}   \label{}
1/ (b_{1,2})_{n+1} =  (\omega \pm i \delta)  -\gamma^2  (b_{1,2})_n \, ,
\end{eqnarray}
where $B_n = \mbox{diag}(b_{1 n}, b_{2 n})$. It is a first-order Riccatti difference equation and can be solved by standard methods:\cite{Kulenovic:2001qd} 
\begin{eqnarray}   \label{retGFnew}
   (b_{1,2})_{n} = \frac{U_{n-1}[(\omega \pm i \delta) /2 \gamma]}{\gamma \, U_{n}[(\omega \pm i \delta) /2 \gamma]} = G^{R,A}_n (\omega)\, ,
\end{eqnarray}
where $U_n(x)$ are the Chebyshev polynomials of the second kind.\cite{szeg1939orthogonal,gogolin2013lectures}  Again, $G^{R,A}_n (\omega)$ are the retarded/advanced GFs of the end sites of the chain. 
Needless to say, this is exactly identical to the result of a straightforward diagonalization of the chain Hamiltonian, see Appendix A.

With Eq.~\eqref{retGFnew} we have achieved more than the conventional approaches: our principal result is the full Keldysh GF, which we find by inverting all transformations,   
\begin{eqnarray}   \label{FfreeChain}
&& F_n(\omega) 
\\ \nonumber
&&= 
 \left(
\begin{array}{cc}
n_F G^R_n + (1-n_F) G^A_n & n_F (G^R_n - G^A_n) \\
(n_F - 1) (G^R_n - G^A_n) & (n_F - 1) G^R_n - n_F G^A_n
\end{array}
\right)
  \, .
\end{eqnarray}
The next recurrence relation we need to solve is
\begin{eqnarray}   \label{}
 \Lambda_{n} = 
-  \Lambda_{n-1}  {F}_{n-1}   \Lambda = (-1)^n  \Lambda \prod_{j=1}^{n-1} ( {F}_{j}  \Lambda ) \, .
\end{eqnarray} 
Using the special properties of the matrices $U$ and $\Lambda$ it can be explicitly computed and found to be
\begin{eqnarray}   \label{LambdafreeChain}
  &&\Lambda_{n+1} = 
(-1)^n  \gamma
\\ \nonumber &&\times
 \left(
\begin{array}{cc}
\frac{n_F}{U_{n}(+)} + \frac{(1-n_F)}{U_n(-)} & - n_F (\frac{1}{U_n(+)} - \frac{1}{U_n(-)}) \\
(1-n_F) (\frac{1}{U_n(+)} - \frac{1}{U_n(-)}) & -\frac{(1-n_F)}{U_n(+)} - \frac{n_F}{U_n(-)}
\end{array}
\right)
  \, ,
\end{eqnarray} 
 whereby $U_n(\pm) = U_{n}[(\omega \pm i \delta) /2 \gamma]$. Plugging the results \eqref{FfreeChain} and \eqref{LambdafreeChain} into equations \eqref{FCS1} 
and \eqref{FCS2} one can derive analytical results for the transport quantities of the chain. 
 As one of the benchmarks for the chain with $N=2$ we find the effective action ${\cal A}'$ [see the definition in Eq.~\eqref{FCS1}] for the non-interacting double quantum dot in sequential geometry,
\begin{widetext}
\begin{eqnarray}
 {\cal A}' = \left(
 \begin{array}{cccc}
  \omega - i \Gamma_L (n_L - 1/2) 
   & 
    i \Gamma_L e^{- i \lambda_L} n_L 
   &
   \gamma
    & 0  \\
   i \Gamma_L e^{i \lambda_L} (n_L - 1)
    & - \omega - i \Gamma_L (n_L - 1/2) & 0 & - \gamma \\
    \gamma 
   & 0 & \omega - i \Gamma_R (n_R - 1/2) & i \Gamma_R e^{- i \lambda_R} n_R \\
 0 & - \gamma & i \Gamma_R e^{i \lambda_R} ( n_R - 1) & - \omega - i \Gamma_R (n_R - 1/2)
\end{array} \right) \, ,
\end{eqnarray}
\end{widetext}
with the help of which one recovers all known results of this setup, also obtainable using the standard techniques, see e.~g. Ref.~[\onlinecite{nazarov2009quantum}].

\begin{figure}
\centering
\includegraphics[width=0.49\textwidth]{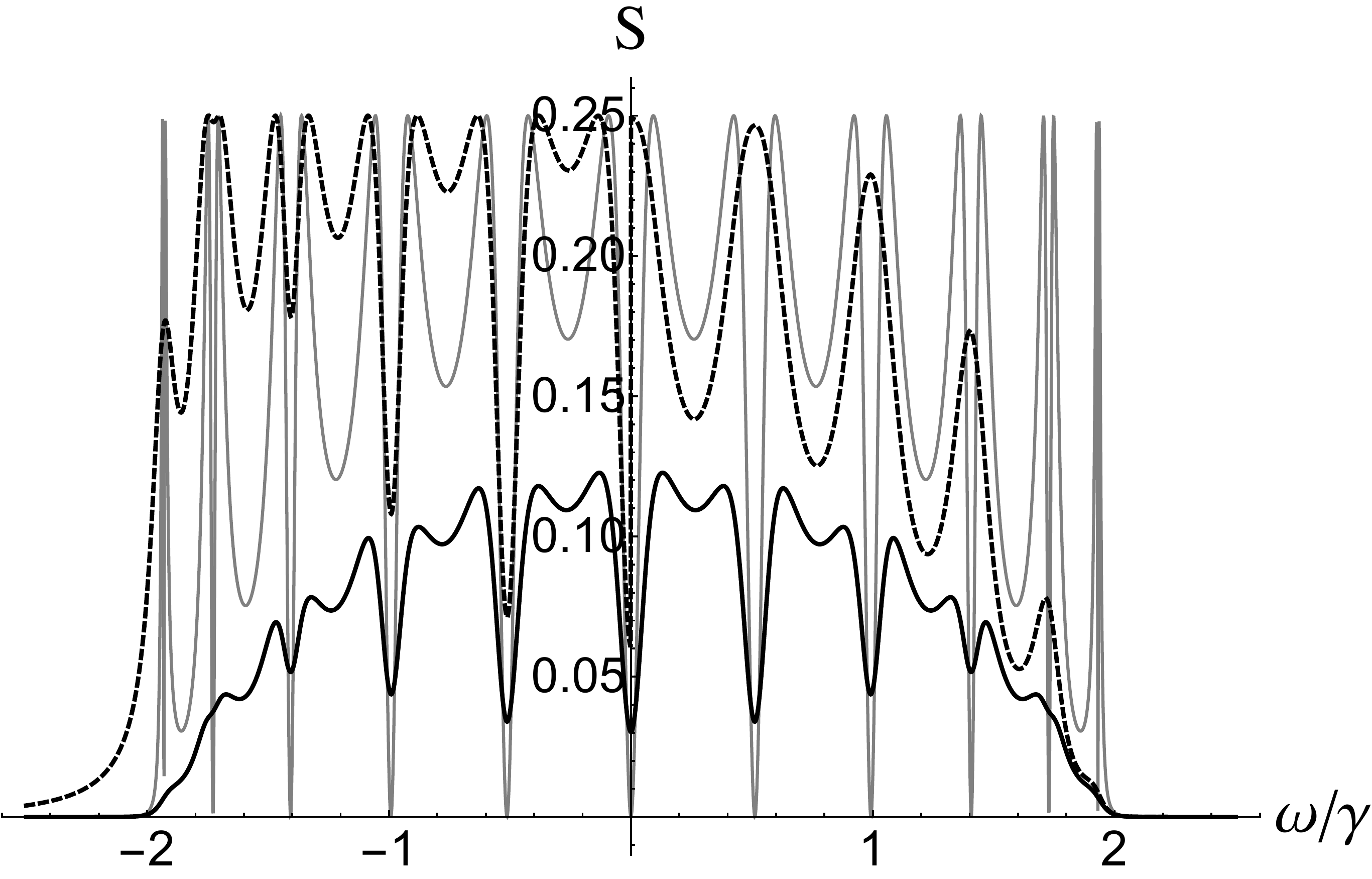}
\caption{  \label{NoiseVSCrossNoiseFreeCase}
Dashed line is the noise $S_R \sim \langle \delta I_R \delta I_R \rangle$ through the right contact and solid line is the cross cumulant  $S_{LR} \sim \langle \delta I_L \delta I_R \rangle$ in a setup with $\Gamma_L=\Gamma_R=\gamma$, $V_L=-V_R=4\gamma$ and $V_G=0$. 
We choose not too long chain $N=10$ and not too small $\delta=0.05 \gamma$ in order to facilitate the comparison with later plots. Grey line represents the same quantities in the case $\delta=0$.  
 }
\end{figure}

From now on we would rather like to concentrate on the current cross correlations (cross cumulants) defined in \eqref{CurrentCross}.
At zero temperature, for $V_L > V_G >V_R$, where $V_G$ is the chemical potential of the substrate, and $\Gamma_L = \Gamma_R = \gamma$, the choice that greatly simplifies expressions but does not affect the physics, and at zero temperature we obtain 
\begin{eqnarray}   \label{}
  S_{LR} = \int \frac{d \omega}{2 \pi} S_{LR}(\omega) \, , 
\end{eqnarray}
with the energy-resolved correlation\footnote{This quantity equals the differential ones considered in Ref.~[\onlinecite{Weithofer2014}] since $S(V) = \partial_V S_{LR}$, when $V_L = V$, $V_R=0$. Considering the energy-resolved correlations allows for arbitrary constellations of applied voltages.}
\begin{eqnarray}   \label{CrossCorr}
 &&S_{LR}(\omega) = (n_L - n_R) 
 \\ \nonumber 
  &&\times
\left| \frac{ U_{N-2}(+)  [1 - U_{N-2}^2(+)/4 - U^2_{N-1}(+)] }{  [1 + U_{N-2}^2(+)/4 - U^2_{N-1}(+) - i U_{N-2}(+) U_{N-1}(+)]^2 } \right|^2 
\end{eqnarray}
In the case of completely insulated wire, when $\delta =0$ it fully coincides with the current correlations $\langle \delta^2 Q_\alpha \rangle/{\cal T}$ locally measured at any of the contacts, see Fig.~\ref{NoiseVSCrossNoiseFreeCase}.
This is not surprising because under such conditions the transport is fully coherent and occurs through single-particle states which extend through the whole wire. As a consequence one can extract the effective transmission coefficient $T(\omega)$ from the energy-resolved current and then compute noise power according to the conventional formula \cite{nazarov2009quantum}
\begin{eqnarray}   \label{}
 S_0 = \int \frac{d \omega}{2 \pi} (n_L - n_R) T(\omega) [1 - T(\omega)]  \, .
\end{eqnarray}
By reversed engineering we obtain a very compact expression for the effective transmission coefficient of the structure, thus far unknown.  

The situation changes completely at finite $\delta \neq 0$. First of all there is an overall suppression of the cross correlation.
A detailed analysis of \eqref{CrossCorr} reveals that it vanishes exponentially $\sim e^{ - N 2 \delta/\gamma}$ for large $N$, for details see Appendix B.
The reason for this effect is rather mundane and can be traced back to the plain currents through the electrodes. They are dominated by the particle out/in flow through the substrate. 
Since in realistic wire-like systems typical $N$ are of the order $10^6$ and larger in order to make $S_{LR}$ perceivable the ratio $\delta/\gamma$ must be extremely small. This limits the cross-correlation observation to small systems, ideally to chains of quantum dots, the length of which should be chosen to be under the critical value $N_c \sim \gamma/\delta$.  

For all $N \gg N_c$ the system can be considered to be infinitely long. In this case the off-diagonal contributions $\Lambda_{N-2}$ and $\Lambda^\dag_{N-2}$ in \eqref{FCS1} vanish and the FCS of the system factorizes into two parts. Each of them describes transport between a half-infinite system and the respective electrode. The GFs for that situation are known and so are the most of the transport properties of such interfaces.

Another striking feature is the asymmetry of the local noise: it is different for $\omega \lessgtr V_G$. The reason for that is quite simple. While for $\omega>V_G$ the particles from the left electrode can flow out of both the right electrode and the ground, for $\omega < V_G$ they can only leave through the right electrode. This leads to higher partition noise in the former case. On the contrary, the cross correlation is completely symmetric.

\subsection{Kitaev chain}
\label{Kitaev_chain}

Kitaev model differs from the tight-binding chain \eqref{H0x} by an increment
\begin{eqnarray}   \label{KitaevD}
 H_1 = \sum_{i=1}^{N-1}  \Delta \left( e^{i \phi} c_{i}^\dag c^\dag_{i+1} +   e^{-i \phi} c_{i+1} c_{i} \right) \, ,
\end{eqnarray}
where $\Delta$ and $\phi$ are the amplitude and phase of intersite triplet superconducting pairing. The full Hamiltonian is given by $H = H_\text{tb} + H_1$ and can also be diagonilized by traditional methods.\cite{Kitaev2001,Alicea2012} We can, however, rederive the known results and analyze the effects of coupling to the environment by the method outlined in the previous Section.  

In equilibrium it is appropriate to work in the Matsubara representation and the chain contraction procedure only needs two-component action composite fields $c_i^\dag =  (c_{i}^\dag, c_{i})$. The inter-site coupling matrix is
\begin{eqnarray}
 \Lambda = \left(
\begin{array}{cc}
 \gamma &  \Delta e^{i \phi} \\
 - \Delta  e^{-i \phi} & - \gamma
\end{array}
\right) \, ,
\end{eqnarray}
and $D_0^{-1} = \mbox{diag}(i \omega_n - \epsilon, i \omega_n+\epsilon)$. Here we continue considering the half-filled case $\epsilon=0$. The simplest quantity we can calculate is the end site GF ${\cal G}_\text{end}(N)$, the components of which are
\begin{eqnarray}
 {\cal G}_{\text{end}} = \left(
\begin{array}{cc}
G & G^+ \\
G^- & \widetilde{G}
 \end{array}
\right) \, , 
\end{eqnarray}
where we define the Matsubara GFs in Nambu representation as
\begin{eqnarray}
G(\tau) = - \langle T_\tau c_1(\tau) c_1^\dag (0) \rangle\, , \, \, \,
G^+(\tau) = - \langle T_\tau c^\dag_1(\tau) c_1^\dag (0) \rangle \, ,    
\nonumber \\ \nonumber 
\widetilde{G}(\tau) = - \langle T_\tau c^\dag _1(\tau) c_1(0) \rangle \, , \, \, \,
G^-(\tau) = - \langle T_\tau c_1(\tau) c_1 (0) \rangle \, . 
\end{eqnarray}
Then in the limit of infinitely long wire $N \to \infty$ we obtain the following equation for the endsite GF [see the definition \eqref{enddef}]:
\begin{eqnarray}
  {\cal G}_{\text{end}}^{-1} = D_0^{-1} -  \Lambda^\dag  {\cal G}_{\text{end}}   \Lambda \, . 
\end{eqnarray}
Its solution is
\begin{widetext}
\begin{eqnarray}   \label{NinftyKitaev}
  G(i\omega_n) &=& \frac{[(i \omega_n)^2 (\Delta^2 + \gamma^2) - 8 \gamma^2 \Delta^2] - 
 (\Delta^2 + \gamma^2) \sqrt{4 \gamma^2 - (i \omega_n)^2} \sqrt{4 \Delta^2 - (i \omega_n)^2}}
 {2 (i\omega_n) (\gamma^2 - \Delta^2)^2} \, ,
 \nonumber \\
 G^+(i\omega_n) &=& - \frac{ \gamma \Delta e^{i \phi} [(i \omega_n)^2 - 2 (\Delta^2 + \gamma^2) - \sqrt{4 \gamma^2 - (i \omega_n)^2} \sqrt{4 \Delta^2 - (i \omega_n)^2}]}
 {(i\omega_n) (\gamma^2 - \Delta^2)^2}  \, .
\end{eqnarray}
\end{widetext}
A bulk site can be considered to be an isolated site simultaneously coupled to two half-infinite chains. That is why for the bulk holds the following relation:
\begin{eqnarray}
  {\cal G}_{\text{bulk}}^{-1} = D_0^{-1} - 2 \Lambda^\dag  {\cal G}_{\text{end}}  \Lambda \, .
\end{eqnarray}
Using these equations one can verify the behaviour of the respective density of states (DOS), showing up signatures of the Majorana fermion states at the ends of the chain. 

One can also analytically solve the system for arbitrary $N$. Unfortunately, the procedure outlined in the previous subsection only works well for the special case $\Lambda^\dag = \Lambda$. We can improve it though. Let us rewrite the recurrence relation for $F_n$ in the form of a Dyson equation,
\begin{eqnarray}        \label{DysonX}
 F_{n+1} = D_0 + D_0 \Lambda^\dag F_n \Lambda F_{n+1} \, ,
\end{eqnarray}
and make the substitution 
\begin{eqnarray}                       \label{SubsT}
F_{n+1} = \Lambda^{-1} P_n P_{n+1}^{-1} \, ,
\end{eqnarray}
then 
\begin{eqnarray}    \label{ChebyshevMatrix}
 P_{n+1} &=& D_0^{-1} \Lambda^{-1} P_n - \Lambda^\dag \Lambda^{-1} P_{n-1}
 \nonumber \\
 &=& A \, P_n - B \, P_{n-1}
  \, 
\end{eqnarray}
is a matrix-valued three-point recurrence relation for a new variable $P_n$. It defines a version of  \emph{Chebyshev matrix polynomials} of the second
kind.\cite{Duran1999304,Zygmunt2002155,FaberTichy2010}\footnote{We would like to remark, that
after a reformulation of the problem as a diagonalization of a large block tridiagonal matrix it is possible to derive a very similar matrix-valued three-point recurrence relation [\onlinecite{Aliaksei2013}]. Although this particular polynomial species does not yield the solutions of \eqref{rec_system}, it is very useful in finding the eigenvalues of the Hamiltonian matrix. }

The solution of \eqref{ChebyshevMatrix} can be constructed employing the standard techniques. Up to a constant we assume it to be of the form $P_r = X^r$ with some yet unknown matrix $X$. Inserting that into the recurrence relation leads to the following quadratic matrix equation,
\begin{eqnarray}    \label{QuadratMatrix}
 X^2 - A \, X + B =0 \, .
\end{eqnarray}
Once the set of its roots $X_i$, $i=1,\dots,m$ is found the solution of the recurrence relation is given by $P_n = \sum_{i=1}^m C_i X_i^n$, where $C_i$ are constants fixed by the initial conditions.\cite{Kulenovic:2001qd} In the equilibrium case of an isolated Kitaev chain one can find the matrices $X_i$ in closed form, see Appendix C. With that one can write down an explicit solution for all necessary  GFs. In the special case $\Delta=0$ and $\delta=0$ in equilibrium all matrices are diagonal and one immediately obtains the solution of the previous subsection. It is also possible to explicitly compute the limit $N \to \infty$ of $F_{N+1} = \Lambda^{-1} P_N P_{N+1}^{-1}$ and to recover the results \eqref{NinftyKitaev}.\cite{Duran1999304} 

We would like to notice, that from the numerical point of view the recurrence relation \eqref{DysonX} is much more efficient than the iteration of Eq.~\eqref{rec_system} as it does not require matrix inversion. Using a more advanced three-point relation \eqref{ChebyshevMatrix} represents yet another speed-up since here one needs less matrix multiplications. The outlined procedure thus offers a higher numerical efficiency even in the absence of analytical solutions.

In the more complicated non-equilibrium case the necessary composite action fields possess 4 components:  $c_i^\dag =  (c_{i -}^\dag, c_{i +}^\dag, c_{i  -}, c_{i  +})$, the kinetic part of the uncoupled system is again simple 
 $D_0^{-1}(\omega) = \mbox{diag}(\omega, - \omega, \omega, - \omega)$ and 
\begin{eqnarray}   \label{}
 \Lambda = \left(
 \begin{array}{cccc}
 \gamma & 0 & \Delta e^{i \phi} & 0 \\
  0 & - \gamma & 0 &- \Delta e^{i \phi} \\
  - \Delta e^{-i \phi}& 0 & - \gamma & 0 \\
  0 & \Delta e^{-i \phi} & 0 & \gamma
 \end{array}
 \right) \, .
\end{eqnarray}
\begin{figure}
\centering
\begin{minipage}{1.\linewidth}
\includegraphics[width=0.9\textwidth]{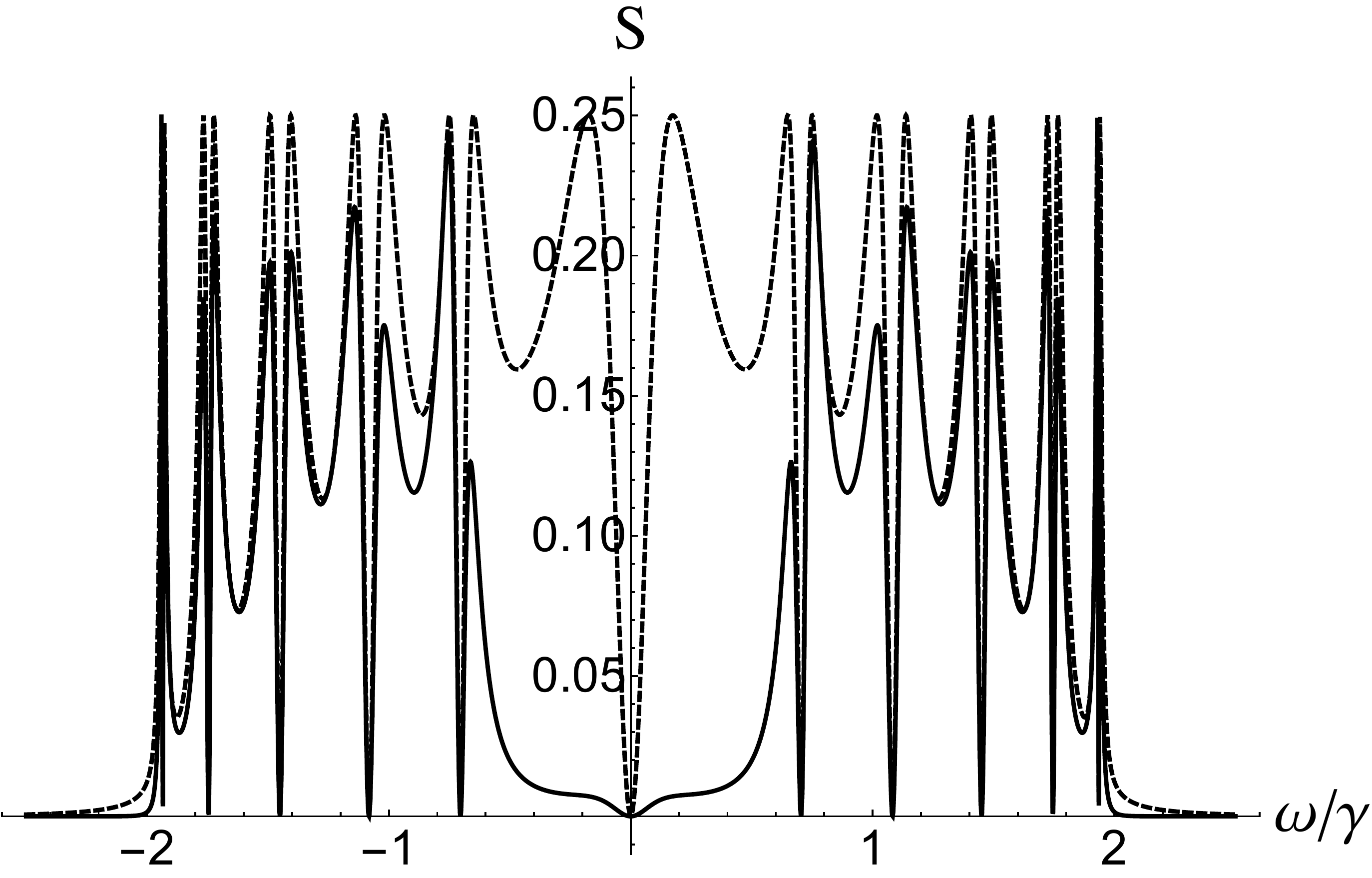}
\end{minipage}
\begin{minipage}{1.\linewidth}
\includegraphics[width=0.9\textwidth]{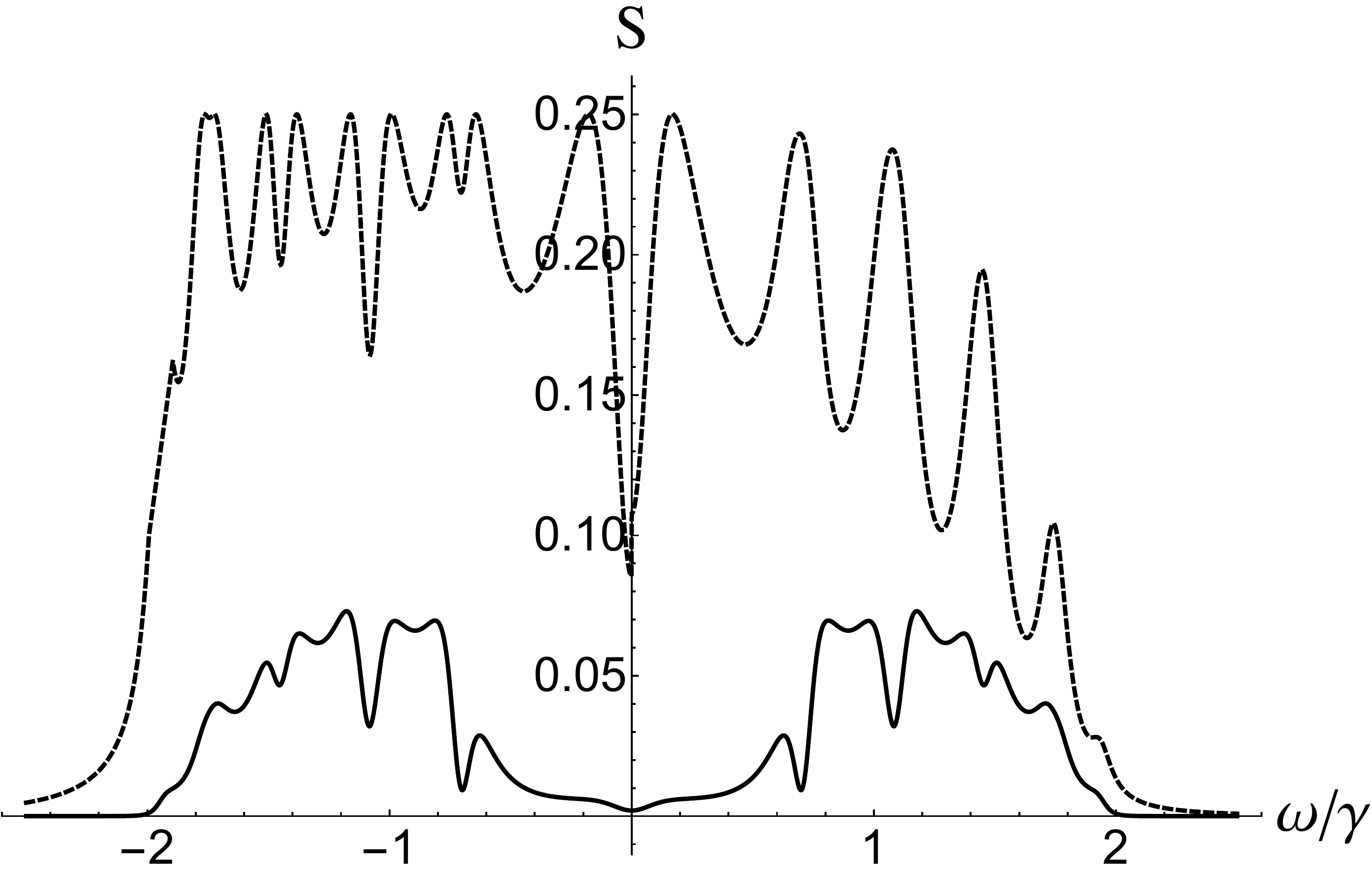}
\end{minipage}
\caption{ \color{black} \label{FigKitaevRNoiseLRNoise}
Dashed line is the noise $S_R \sim \langle \delta I_R \delta I_R \rangle$ through the right contact and solid line is the cross cumulant  $S_{LR} \sim \langle \delta I_L \delta I_R \rangle$ in a setup with $\Gamma_L=\Gamma_R=\gamma/2$, $\Delta=\gamma/4$, $V_L=-V_R=4\gamma$, $V_G=0$. 
(Top) $\delta=0$, 
 (Bottom) $\delta=0.025\gamma$. We choose not too long chain $N=10$ and not too small $\delta$ in order to make the suppression of cross correlations visible.  }
\end{figure}

The FCS of the system can be found using Eq. \eqref{FCS2}. The not grounded situation $\delta=0$ has been considered in [\onlinecite{Weithofer2014}]. We recover all previous results that is why we would like to concentrate on the situation of finite $\delta \neq 0$. Here an analytical solution does not exist and we resort to numerical calculation using the highly efficient chain contraction procedure 
yielding \eqref{FCS1}. 

In Fig.~\ref{FigKitaevRNoiseLRNoise} we plot the energy-resolved noise on the right contact as well as the cross correlation. At $\omega=0$, which is the position of the Majorana edge states the transmission is perfect. The corresponding conductivity reproduces the form of the edge state which is due to the coupling of the electrodes a Lorentzian with the width depending on $\Gamma_{L,R}$. Accordingly both the noise and the cross correlations approach zero. Beyond that central feature already for the isolated chain $\delta=0$ the cross correlation is considerably suppressed as compared to the local noise. $S_{LR}$ is much weaker in the subgap regime as here the effective bulk transmittance is small. It is known to be exponential in chain length and the gap parameter $\sim e^{- N \Delta/\gamma}$. 
This is in strong contrast to the same system in normal state, in which both correlations are equal to each other, see Section \ref{Nonintchain}. 

\begin{figure}
\centering
\includegraphics[width=0.45\textwidth]{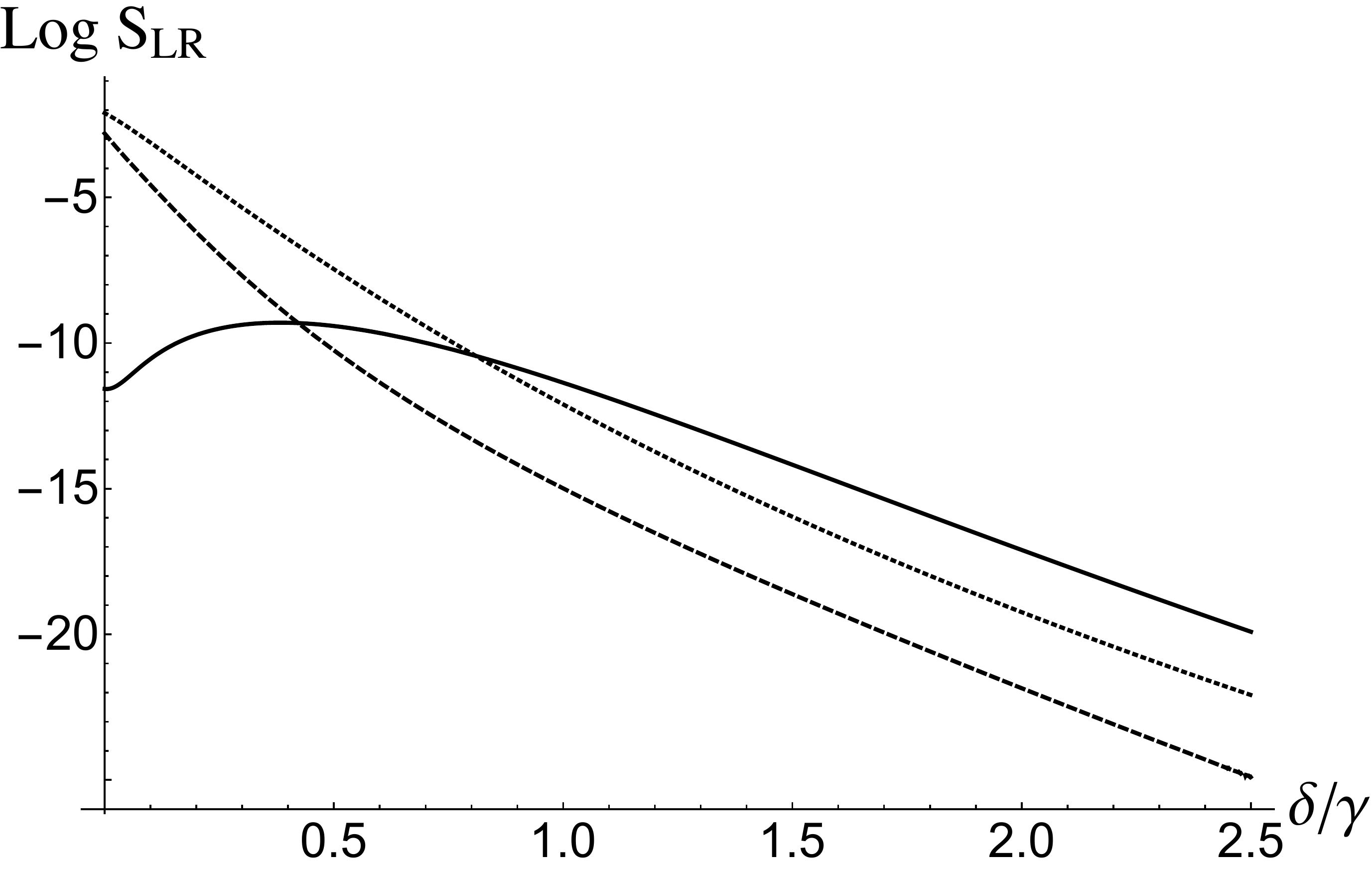}
\caption{  \label{LogSLRDifferentdelta}
Logarithmic plot of the cross correlation $S_{LR}$ for $\Delta=\gamma/2$, within the gap at $\omega=0.1\gamma$ (solid line) and outside of the gap at $\omega=1.2 \gamma$ (dashed line) as a function of the coupling strength $\delta$. Dotted line represents the result for the chain in the normal state $\Delta=0$. Other parameters are as in Fig.~\ref{FigKitaevRNoiseLRNoise}: $\Gamma_L=\Gamma_R=\gamma/2$, $V_L=-V_R=4\gamma$, $V_G=0$. 
 }
\end{figure}

In the situation with $\delta \neq 0$ there is an exponential suppression of the cross correlation also outside of the gap region. Here the suppression factor is $\sim e^{- N f(\Delta,\delta)/\gamma}$, where $f(\Delta,\delta)$ is a complicated function with the asymptotics lim${}_{\Delta \to 0} f(\Delta,\delta) \to \delta$, see Fig.~\ref{LogSLRDifferentdelta}.
Furthermore, the local noise is asymmetric with respect to the ground chemical potential for the same reasons as in the free case $\Delta=0$, see previous subsection.

\begin{figure}
\centering
\begin{minipage}{0.9\linewidth}
\includegraphics[width=1.\textwidth]{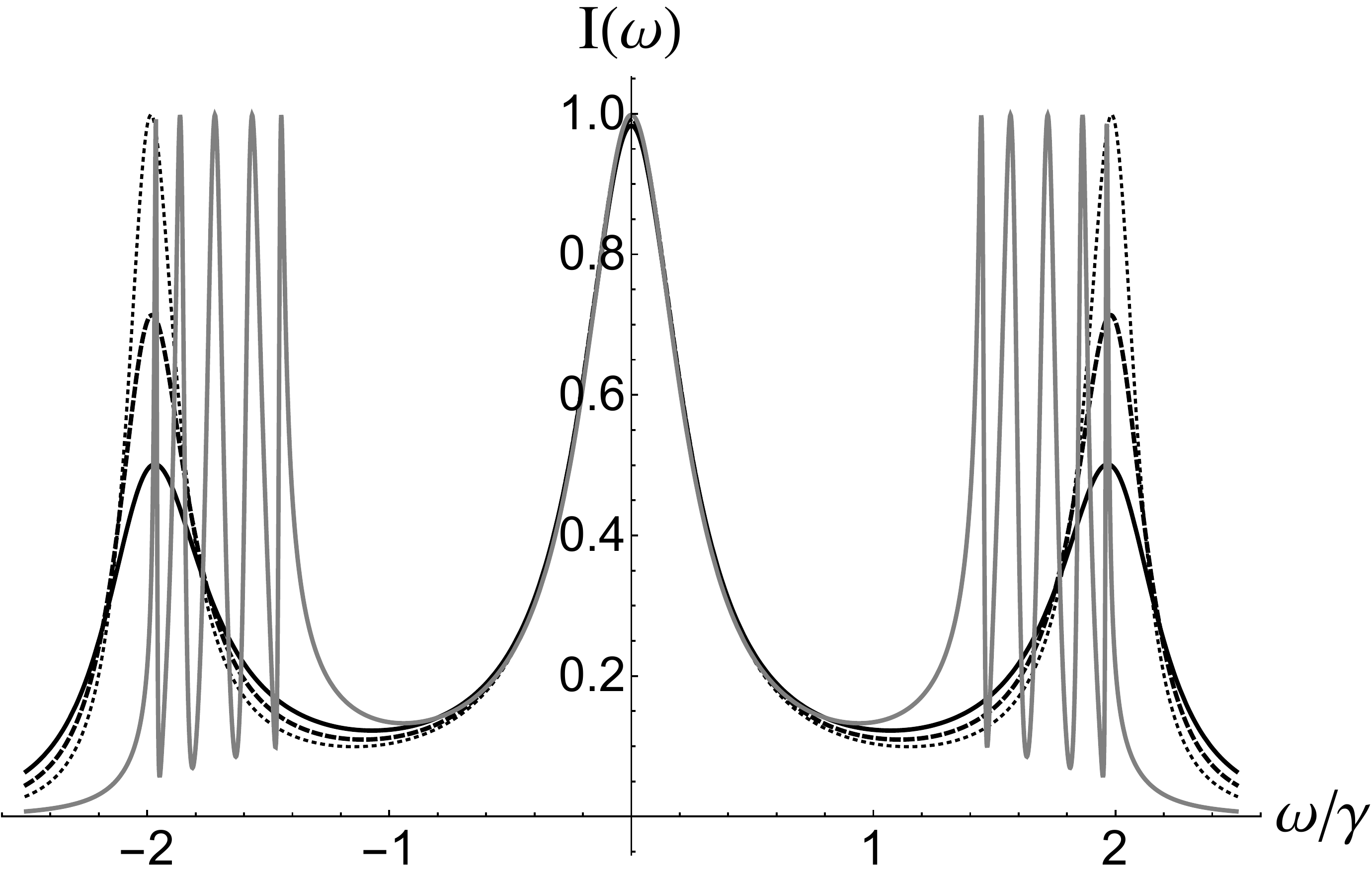}
\end{minipage}
\begin{minipage}{0.9\linewidth}
\includegraphics[width=1.\textwidth]{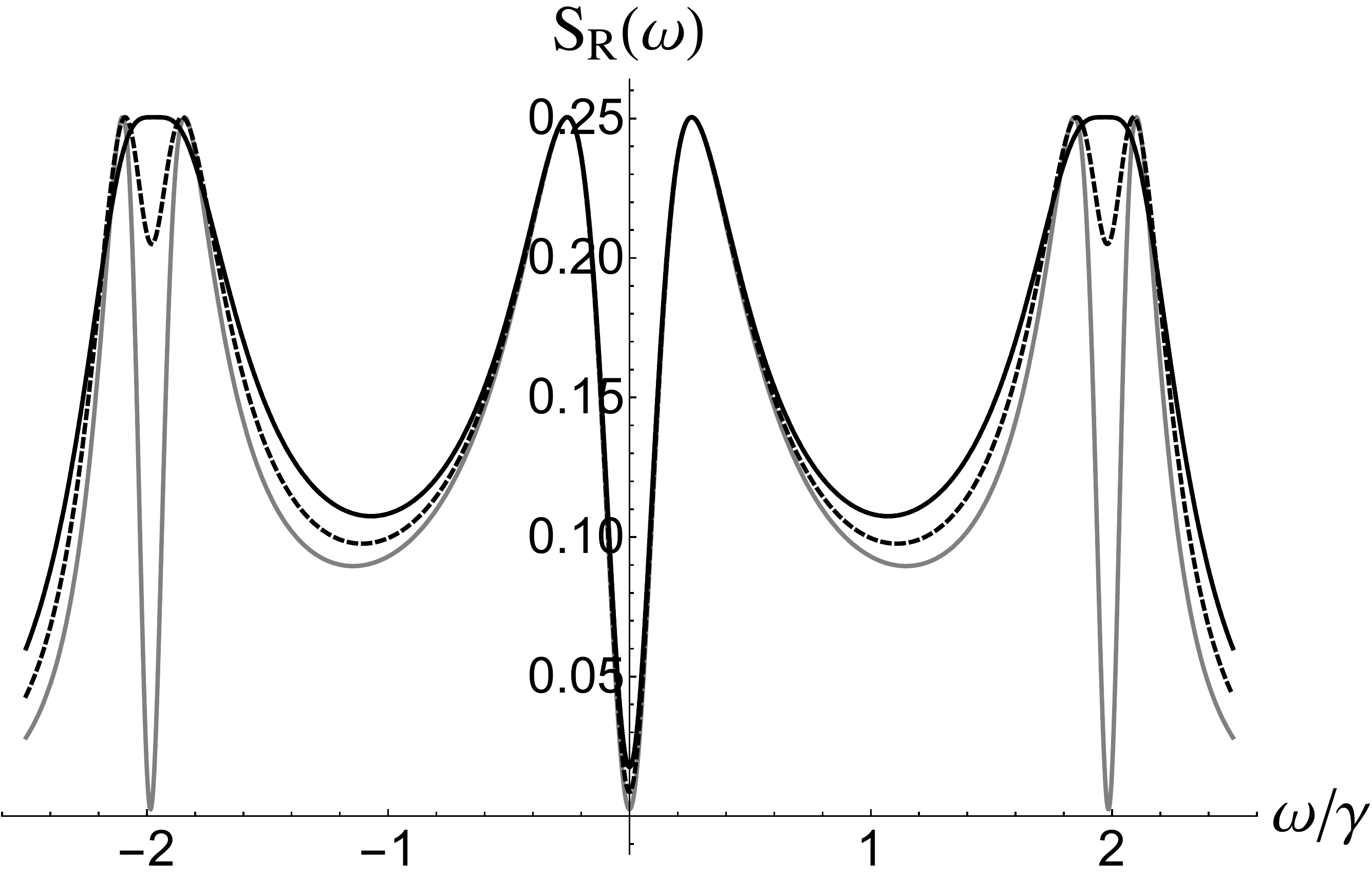}
\end{minipage}
\caption{ \color{black} \label{IVresonance}
(Top) Energy-resolved current measured locally for $\Gamma_L=\Gamma_R=\gamma/2$, $V_G=0$ and $V_L=-V_R=4\gamma$. Dotted line: $N=56$ and $\delta=0$, dashed line: $N=56$ and $\delta=0.1\gamma$, solid line: $N=2$ and $\delta=0.1\gamma$, all of them at resonance $\gamma=\Delta$. For comparison we plot the data for the off-resonant system at $\Delta=0.7\gamma$ and $N=12$ (grey line).   (Bottom)
Energy-resolved noise for different chain lengths $N=56$: grey line for $\delta=0$, dashed line for $\delta=0.1\gamma$; and $N=2$: solid line represents the data for a range of $0<\delta<\gamma$. All other parameters are as in the upper panel. 
}
\end{figure}

At the point $\gamma=\Delta$ when the gap equals the bandwidth the situation changes profoundly. Here not only the plain energy-resolved current but also the local noise ceases to depend on the chain length, see Fig.~\ref{IVresonance}. It is best seen in the data for chains with lengths $N=56$ and $N=2$, which differ only slightly and only in vicinity of the band borders. In effect the overall energy dependence loses its structure as if a chain of any length is indeed equivalent to two sites only. This is in accordance with the picture in which at resonance $\gamma=\Delta$ the end Majoranas are effectively detached from the system. \cite{Kitaev2001,Weithofer2014}
The fate of the cross correlations is even more spectacular as they completely vanish exactly at resonance. This puts a bound on the applicability of the effective Hamiltonian $H_M$ proposed in [\onlinecite{GolubHorovitz2011}]. In fact, our method allows for an explicit derivation of the improved version of $H_M$.

\subsection{TB chain with spin-orbit coupling and proximity-induced BCS pairing}

One of the successful realizations of the Kitaev model is based on a semiconductor nanowire with strong spin-orbit interaction (SOI), which is in immediate contact with a conventional BCS superconductor. It induces singlet pairing via proximity effect. Aided by the SOI the system can be brought into a topologically non-trivial ground state with the properties similar to those of the Kitaev chain.\cite{OregRefaelvonOppen2010,Lutchyn2010,PhysRevB.90.235415} 

The action for such a system is best formulated in the language of the composite fields \eqref{LongC}. For the on-site contribution we have
\begin{eqnarray}
 D_0(\omega) = 
 \left(
\begin{array}{cc}
 D_\uparrow(\omega) & K_\Delta \\
 K_\Delta &  D_\downarrow(\omega)
\end{array}
\right) \, ,
\end{eqnarray}
where 
\begin{eqnarray}
  D^{-1}_\sigma(\omega) = \mbox{diag}\left[ \omega - (\epsilon + \sigma h),  -\omega + (\epsilon + \sigma h), 
  \right. \nonumber \\ \left.
  \omega + (\epsilon + \sigma h), -\omega - (\epsilon + \sigma h)\right] \, .
\end{eqnarray}
Here $h = \mu_B g B$ is the applied magnetic field in energy units, $\mu_B$ is the Bohr's magneton and $g$ Land\'{e} factor of the electron. The BCS correlations are described by
\begin{eqnarray}
 K_\Delta = \left(
\begin{array}{cccc}
 0 & 0 & \Delta e^{i \phi} & 0 \\
 0 & 0 & 0 & -\Delta e^{i \phi} \\
 - \Delta e^{-i \phi} & 0 & 0 & 0 \\
 0 & \Delta e^{-i \phi} & 0 & 0
\end{array}
\right) \, ,
\end{eqnarray}
with $\Delta$ being the order parameter and $\phi$ the SC phase. The inter-site coupling is mediated by the following matrix elements:
\begin{eqnarray}
 \Lambda = \left(
\begin{array}{cc}
  \Lambda_\gamma &  \Lambda_\alpha \\
  -  \Lambda_\alpha &   \Lambda_\gamma
\end{array}
\right) \, , 
\end{eqnarray}
whereby $\Lambda_\gamma = \mbox{diag}(\gamma, - \gamma, - \gamma, \gamma)$ and $\Lambda_\alpha = \mbox{diag}(\alpha, - \alpha, - \alpha, \alpha)$. $\alpha$ is the strength of the SOI in energy units. 

As in the case of the Kitaev chain, an explicit analytic solution of the relations \eqref{rec_system} is unfortunately not possible and one needs to resort to numerics. Thus far a large number of works discussed the conductance properties of such systems, see e.~g.~[\onlinecite{OregRefaelvonOppen2010,PhysRevB.85.144525,PhysRevB.87.024515}]. Here we would like to concentrate on the effects of the environment. It turns out that as expected most features of the transport characteristics coincide with those of the Kitaev chain even in their energy-resolved version. For instance, both the plain current as well as the local and non-local correlations are asymmetric with respect to zero energy as soon as a finite $\delta$ coupling to the environment is introduced. Similar to the Kitaev chain, the cross correlation decays exponentially in the chain length and $\delta$. This suppression factor
has the generic form $\sim \exp[- \delta N f(\Delta, \alpha, h)/\gamma]$. The $\Delta$ dependence essentially follows that for the Kitaev chain. $f(\alpha,h)$ grows linearly with the magnetic field  up to the value of the order of the bandwidth $h \sim 2 \gamma$. Beyond that point it diverges rapidly enforcing a vanishing cross-correlation. This scenario holds for all $\delta$. On the contrary, $\alpha$ affects $S_{LR}$ in a very non-universal way irrespectively of the coupling strength to the environment.

\section{Conclusions}

We have developed a general strategy to compute the cumulant generating function of charge transport through quantum wires of finite length coupled to external electrodes. This chain contraction procedure is based on recurrence relation technique for the Green's functions and represents its extension to non-equilibrium situations. Furthermore, we show how the emergent recurrence relations can be analytically solved with the help of matrix polynomials of Chebyshev type. Using this approach we analyze the behaviour of the lowest order cumulants of charge transport through a wire coupled to an environment via particle exchange. We show that the cross correlation between the currents through the source and drain contacts decays exponentially in the strength of the coupling to the environment. This picture remains robust also in regimes in which topologically non-trivial Majorana states emerge at the ends of the wire. One avenue for further research could be the extension of the developed techniques to systems in preparational non-equilibrium, the typical realization of which is e.~g. a tunnelling quench, to chains of interacting quantum dots, or to local conductance properties.\cite{TransientMajorana,Chinese2006,NonEqAIM,DoornenbalStoof2015}

\acknowledgements

The author is supported by the Heisenberg Programme of the Deutsche Forschungsgemeinschaft (Germany) under Grant No. KO 2235/5-1.


\section*{Appendix A}   \label{AppA}

Here we demonstrate how one can recover the results of the conventional approaches from the solutions of recurrence relations. The simplest quantity is the retarded GF given in Eq.~\eqref{retGFnew}. It is well known that its poles yield the dispersion relation of the excitations in the system. Using the trigonometric representation of the Chebyshev polynomials\cite{gradshteyn2007} 
\begin{eqnarray}   \label{}   \nonumber
 U_N(x) = \frac{\sin \left[ (N+1) \mbox{arccos} \, x \right]}{\sin (\mbox{arccos} \, x)} \, ,
\end{eqnarray}
we can immediately write down the corresponding equation for the dispersion and find it to be given by 
\begin{eqnarray}   \label{dis1}
 E_j = 2 \gamma \cos \left( \frac{2 \pi j}{N+1} \right) \, , \, \, \, \, \, j=-N, \dots, N \, ,
\end{eqnarray}
which is an expression known from the direct diagonalization of the tight-binding Hamiltonian. 
Furthermore, by using the explicit form of the single-particle wave functions one can also recover the GF mentioned above in its trigonometric representation. 

By expansion of \eqref{dis1} for small $j/N$ one then reproduces the dispersion of free particles. A system with discrete translational symmetry and continuous $k$ then emerges for $2 \pi j/(N+1) \to k$, where $k$ is the wave vector restricted to the first Brillouine zone $-\pi < k < \pi$, 
\begin{eqnarray}   \label{Edisp}
 E(k) = 2 \gamma \, \cos \, k \, .
\end{eqnarray}
 This is appropriate for systems with $N \to \infty$. This particular limit can be safely reproduced from the recurrence relation solution. 
A direct computation of a quotient of two polynomials in the limit $N \to \infty$  is not trivial but possible and one gets (a mathematically rigorous procedure is discussed in Ref.~[\onlinecite{Duran1999304}]) 
\begin{eqnarray}   \label{Sol1}
 G_\infty(i \omega_m) &=& \frac{1}{2 \gamma^2 G_0(i \omega_m)} - \sqrt{\frac{1}{[2 \gamma^2 G_0(i \omega_m)]^2} - \frac{1}{\gamma^2}} 
 \nonumber \\
 &=& \frac{1}{2 \gamma^2} \left( i \omega_m - \sqrt{(i \omega_m)^2 - 4 \gamma^2} \right) \, .
\end{eqnarray}
Exactly the same result is obtained in a somewhat sloppily way by taking the limit $N \to \infty$ in the recurrence relation for the retarded GF at the end site:
\begin{eqnarray}   \label{Dyson1}
 G^{-1}_{N+1}(i \omega_m) = G^{-1}_0(i \omega_m) - \gamma^2 G_N(i \omega_m) \, , 
\end{eqnarray}
for $G_{N+1}=G_{N}=G_\infty(i \omega_m)$. Eq.~\eqref{Dyson1} then leads to a quadratic equation, the solution of which is precisely the expression \eqref{Sol1}. 
 GF for a bulk site can easily be computed by observing that the correct self-energy for a site in the bulk is equal to a doubled GF of an end site of a half-infinite system. Thus we obtain the identity 
\begin{eqnarray}   \label{bulkGF}
 G_{\text{bulk}}(i \omega_m) &=& \left[
  G^{-1}_0(i \omega_m) - 2 \gamma^2 G_\infty(i \omega_m)
 \right]^{-1} 
 \nonumber \\
 &=&
 1/ \sqrt{(i \omega_m)^2 - 4 \gamma^2}
 \, .
\end{eqnarray}
On the other hand, these local GFs can be computed using continuum $k$ taking into account that the wave functions of single particles are: (i) $\psi(x) \sim \sin( k x )$ for a system with open boundaries and: (ii) $\psi(x) \sim e^{i k x}$ in an infinitely large system. From the spectral representation we have
\begin{eqnarray}     \label{Cont0}
 G_x(i \omega_m) = \int_0^\pi \frac{d k}{2 \pi} \frac{\psi^2(k x)}{i \omega_m - E(k)} \, .
\end{eqnarray}
For the end site, $x=1$ and in the case (i) the result is compatible with \eqref{Sol1}.  In the case (ii), when all points $x$ are in the bulk, the integrand does not depend on $x$ any more and we recover Eq.~\eqref{bulkGF}.

\section*{Appendix B}   \label{AppB}

The derivation of the length dependence of the cross correlation is easiest for small voltages, where only $\omega \to 0$ matters. The arguments of the Chebyshev polynomials entering \eqref{CrossCorr} are then just $x=\pm i \delta/\gamma$. We start with the standard representation in the form \cite{gradshteyn2007}
\begin{eqnarray}   \label{} \nonumber
 U_N(x) &=& \frac{1}{2 i \sqrt{1 - x^2}} \left[ 
  (x + i \sqrt{1 - x^2})^{N+1} 
  \right. \\ \nonumber 
  &-& \left. ( x - i \sqrt{1 - x^2})^{N+1}
 \right] \, .
\end{eqnarray}
Inserting $x= i \delta/\gamma$ and keeping the terms of leading order only we obtain (we recall that $\delta$ is always positive)
\begin{eqnarray}   \label{} 
&& U_N( i \delta/\gamma) = \frac{i^N}{2} \left[ ( 1 + \delta/\gamma)^{N+1} 
 - (-1)^{N+1}
 \nonumber \right. \\ \nonumber
 &\times& \left.  ( 1 - \delta/\gamma)^{N+1} \right] 
 \approx \frac{i^N}{2} \left[ e^{\delta (N+1)/\gamma)} 
 \right. \\ \label{asy1} \nonumber
 &-& \left.  (-1)^{N+1} e^{- \delta(N+1)/\gamma} \right]  \approx \frac{i^N}{2} e^{\delta (N+1)/\gamma} \, .
\end{eqnarray}
In the similar way one can derive
 $U_N( -i \delta/\gamma) \approx (-i)^N e^{\delta (N+1)/\gamma}/2$. 
Inserting the last two results into Eq.~\eqref{CrossCorr} and taking the limit of large $N$ one then obtains the asymptotics given in the main text.

\section*{Appendix C}   \label{AppC}

In order to simplify we notation we use $\omega$ instead of the Matsubara frequencies $i \omega_n$. Then the matrix equation \eqref{QuadratMatrix} has $m=6$ different solutions: 
\begin{eqnarray}   \label{} \nonumber
 X = \left(
\begin{array}{cc}
 x_1 & x_2 \\
 x_3 & x_4
\end{array}
\right) \, ,
\end{eqnarray}
with
\begin{eqnarray}   \label{}
x_1 = &&\frac{\gamma \left[ \omega^2 + 4 \Delta^2 \pm \sqrt{(\omega^2 - 4 \gamma^2)(\omega^2 - 4 \Delta^2)}\right]}{2 \omega (\gamma^2 - \Delta^2)} \, ,
\nonumber \\
&&\frac{ \omega (\gamma^2 + \Delta^2) \pm \gamma \sqrt{(\gamma^2 -  \Delta^2)(\omega^2 - 4 \gamma^2)}}{2 \gamma (\gamma^2 - \Delta^2)} \, ,
\nonumber \\ \nonumber
&&\frac{ 2 \omega \gamma \pm \gamma \sqrt{(\gamma^2 -  \Delta^2)(\omega^2 - 4 \Delta^2)}}{2  (\gamma^2 - \Delta^2)}\, . 
\end{eqnarray}
\begin{eqnarray}   \label{}
x_2 = 
&&-\frac{\Delta \left[ \omega^2 + 4 \gamma^2 \pm \sqrt{(\omega^2 - 4 \gamma^2)(\omega^2 - 4 \Delta^2)}\right]}{2 \omega (\gamma^2 - \Delta^2)} \, ,
\nonumber \\ \nonumber
&&-\frac{ \Delta \left[ 2\gamma \omega  \pm \sqrt{(\gamma^2 -  \Delta^2)(\omega^2 - 4 \gamma^2)}\right]}{2 \gamma (\gamma^2 - \Delta^2)} \, ,
\nonumber \\ \nonumber
&&-\frac{  \omega (\gamma^2 + \Delta^2) \pm \gamma \sqrt{(\gamma^2 -  \Delta^2)(\omega^2 - 4 \gamma^2)}}{2 \Delta (\gamma^2 - \Delta^2)}\, . 
\end{eqnarray}
\begin{eqnarray}   \label{} \nonumber
x_3 = 
&&\frac{\Delta \left[ \omega^2 + 4 \gamma^2 \pm \sqrt{(\omega^2 - 4 \gamma^2)(\omega^2 - 4 \Delta^2)}\right]}{2 \omega (\gamma^2 - \Delta^2)} \, ,
\nonumber \\
&&\frac{ \Delta \left[ 2\gamma \omega  \mp \sqrt{(\gamma^2 -  \Delta^2)(\omega^2 - 4 \gamma^2)}\right]}{2 \gamma (\gamma^2 - \Delta^2)} \, ,
\nonumber \\ \nonumber
&&\frac{  \omega (\gamma^2 + \Delta^2) \mp \gamma \sqrt{(\gamma^2 -  \Delta^2)(\omega^2 - 4 \gamma^2)}}{2 \Delta (\gamma^2 - \Delta^2)}\, . 
\end{eqnarray}
\begin{eqnarray}   \label{}  \nonumber
x_4 = &&-\frac{\gamma \left[ \omega^2 + 4 \Delta^2 \pm \sqrt{(\omega^2 - 4 \gamma^2)(\omega^2 - 4 \Delta^2)}\right]}{2 \omega (\gamma^2 - \Delta^2)} \, ,
\nonumber \\
&&-\frac{ \omega (\gamma^2 + \Delta^2) \mp \gamma \sqrt{(\gamma^2 -  \Delta^2)(\omega^2 - 4 \gamma^2)}}{2 \gamma (\gamma^2 - \Delta^2)} \, ,
\nonumber \\ \nonumber
&&-\frac{ 2 \omega \gamma \mp \gamma \sqrt{(\gamma^2 -  \Delta^2)(\omega^2 - 4 \Delta^2)}}{2  (\gamma^2 - \Delta^2)}\, . 
\end{eqnarray}

\bibliographystyle{apsrev}
\bibliography{Long_Dot_Chains.bib}

\end{document}